\documentstyle[emulateapj,onecolfloat]{article}
\newcommand{\bn}{\hat{\bf n}}
\newcommand{\bl}{{\bf l}}
\newcommand{\bll}{{\bf L}}

\newcommand{\ApJ}{ApJ}
\newcommand{\ApJL}{ApJ Lett.}
\newcommand{\PRL}{Phys. Rev. Lett.}
\newcommand{\PRD}{Phys. Rev. D}

\newcommand{\etal}{et al.}
\newcommand{\ARAA}{ARA\&A}
\newcommand{\AsAs}{A\&A}
\newcommand{\amp}{\& }
\newcommand{\aut}[2]{{#1, #2.}}
\newcommand{\laut}[2]{{#1, #2.}}
\newcommand{\refs}[6]{#5, #2, #3,  {#4}.}

\newcommand{\mybib}[2]{\bibitem[#1]{#2}}
\newcommand{\intl}[1]{\int {d^2 l_{#1} \over (2\pi)^2}}
\newcommand{\intlp}[1]{\int {d^2 l_{#1}' \over (2\pi)^2}}
\newcommand{\vsp}{\vphantom{\Big[}\\}
\newcommand{\dotfac}[1]{({\bll} \cdot {\bf l}_{#1})}
\newcommand{\lsim}{\lesssim}
\newcommand{\gsim}{\gtrsim}
\begin{document}
\twocolumn[
\title{Mass Reconstruction with CMB Polarization}
\author{Wayne Hu$^{1,2}$ and Takemi Okamoto$^3$}
\affil{${}^{1}$Center for Cosmological Physics, University of Chicago, Chicago, IL 60637\\
${}^{2}$Department of Astronomy and Astrophysics and Enrico Fermi Institute,\\
University of Chicago, Chicago, IL 60637\\
${}^{3}$Department of Physics, University of Chicago, Chicago, IL 60637\\
}

\begin{abstract}
Weak gravitational lensing by the intervening large-scale 
structure of the Universe induces high-order correlations 
in the cosmic microwave background (CMB) temperature and polarization
fields.  We construct minimum variance estimators
of the intervening mass distribution out of the six 
quadratic combinations of the temperature and polarization fields.  
Polarization begins to assist in the reconstruction when $E$-mode 
mapping becomes possible on degree-scale fields, i.e. for an experiment 
with a noise level of $\sim 40 \mu$K-arcmin
and beam of $\sim 7'$, similar to the Planck
experiment; surpasses the temperature reconstruction at $\sim 26 \mu$K-arcmin
and $4'$; yet continues to improve the reconstruction until the
lensing $B$-modes are mapped to  $l \sim 2000$ at 
$\sim 0.3 \mu$K-arcmin and $3'$.  Ultimately, the correlation between the 
$E$ and $B$ modes can provide a high signal-to-noise mass map out 
to multipoles of $L \sim 1000$, extending the range of 
temperature-based estimators by nearly an order of magnitude.
We outline four applications of mass reconstruction: measurement of 
the linear power spectrum in projection to the cosmic variance limit 
out to $L \sim 1000$ (or wavenumbers $0.002 \lsim k \lsim 0.2$ in $h$/Mpc), 
cross-correlation with cosmic shear surveys to probe 
the evolution of structure tomographically, cross-correlation of 
the mass and temperature maps to probe the dark energy, and the 
separation of lensing and gravitational wave $B$-modes.
\end{abstract}

]

\section{Introduction}

The weak gravitational lensing of cosmic microwave background (CMB) temperature and polarization
anisotropies provides a unique opportunity to map the distribution of
matter on large scales and high redshift where density fluctuations were
still linear.   Although lensing effects are apparent in the power 
spectra of temperature and polarization (\cite{Sel96} 1996; \cite{ZalSel98}
1998), it is the higher order correlations induced by lensing that make
mass reconstruction possible (\cite{Ber97} 1997).  

By remapping the CMB fields according to potential gradients, 
lensing acts as a convolution in Fourier space which
introduces correlations between angular wavenumbers or multipole moments.  
From a quadratic
combination of the multipoles, one can form estimators of the
potential field and hence the intervening mass.  
\cite{ZalSel99} (1999) and \cite{GuzSelZal00} (2000) constructed 
noisy estimators out of the product of gradients of the temperature and
polarization fields.  \cite{Hu01} (2001a,b) showed that the minimum
variance estimator constructed from the temperature field has substantially
greater signal-to-noise with arcminute resolution CMB maps.  This 
estimator
enables mapping of the dark matter above the degree scale, 
where the deflection power peaks.  
The cosmic variance of the CMB temperature field 
itself prevents mapping on smaller scales.
In this Paper, we show how this limitation can be overcome with
minimum variance quadratic estimators involving the polarization field.

The CMB polarization field in principle provides a more direct probe of
lensing than the temperature field.  
Unlike the temperature anisotropy,
there is negligible cosmological contamination of the polarization
field in the arcminute regime (\cite{Hu00a} 2000a). 
Furthermore, density perturbations in the
linear regime generate only the so-called $E$-mode polarization 
(\cite{KamKosSte97} 1997;
\cite{ZalSel97} 1997).  Lensing converts $E$-mode
polarization to its complement, the $B$-mode polarization (\cite{ZalSel98} 1998).  
Although gravitational waves also generate $B$-mode polarization,
they do so only above the degree scale.  
\cite{Hu00b} (2000b) and \cite{BenBervan01} (2001) used the induced
correlation between the $E$ and the $B$-modes to construct statistical
measures of the lensing.  Here we show that the correlation allows
a direct reconstruction of the lensing masses which in fact has in 
principle the highest signal-to-noise of all the quadratic estimators.

In practice, achieving this potential in the presence of detector noise, systematics
and foreground contamination of the polarization will be challenging.
These same challenges will also have to be overcome in order to
probe the physics of the early universe through gravitational
wave $B$-mode polarization. A lensing study can therefore 
be conducted as secondary science for an experiment
devoted to gravitational waves.   In fact, as the leading
cosmological contaminant of the gravitational wave $B$-modes,
a lensing study may well be required of such an experiment
(\cite{Hu01c} 2001c).

We begin in \S \ref{sec:lensing} with a brief review of lensing
effects on the temperature and polarization fields.  In \S \ref{sec:minvar},
we present a formal study of the
minimum variance quadratic estimators of the lensing potential and show that
the $EB$ combination can produce a high signal-to-noise
mass map out to the $10'$ scale.  We explicitly construct this
estimator in \S \ref{sec:eb} and simulate its performance in the
presence of detector noise.   We discuss applications of
mass reconstruction in \S \ref{sec:applications}.
For illustrative purposes, we use a flat $\Lambda$CDM cosmology throughout 
with parameters
$\Omega_c = 0.3$, $\Omega_b=0.05$, $\Omega_\Lambda=0.65$, $h=0.65$, $n=1$,
$\delta_H=4.2\times 10^{-5}$ and no gravitational waves.

\begin{figure*}[t]
\centerline{
\epsfxsize=2.25truein\epsffile{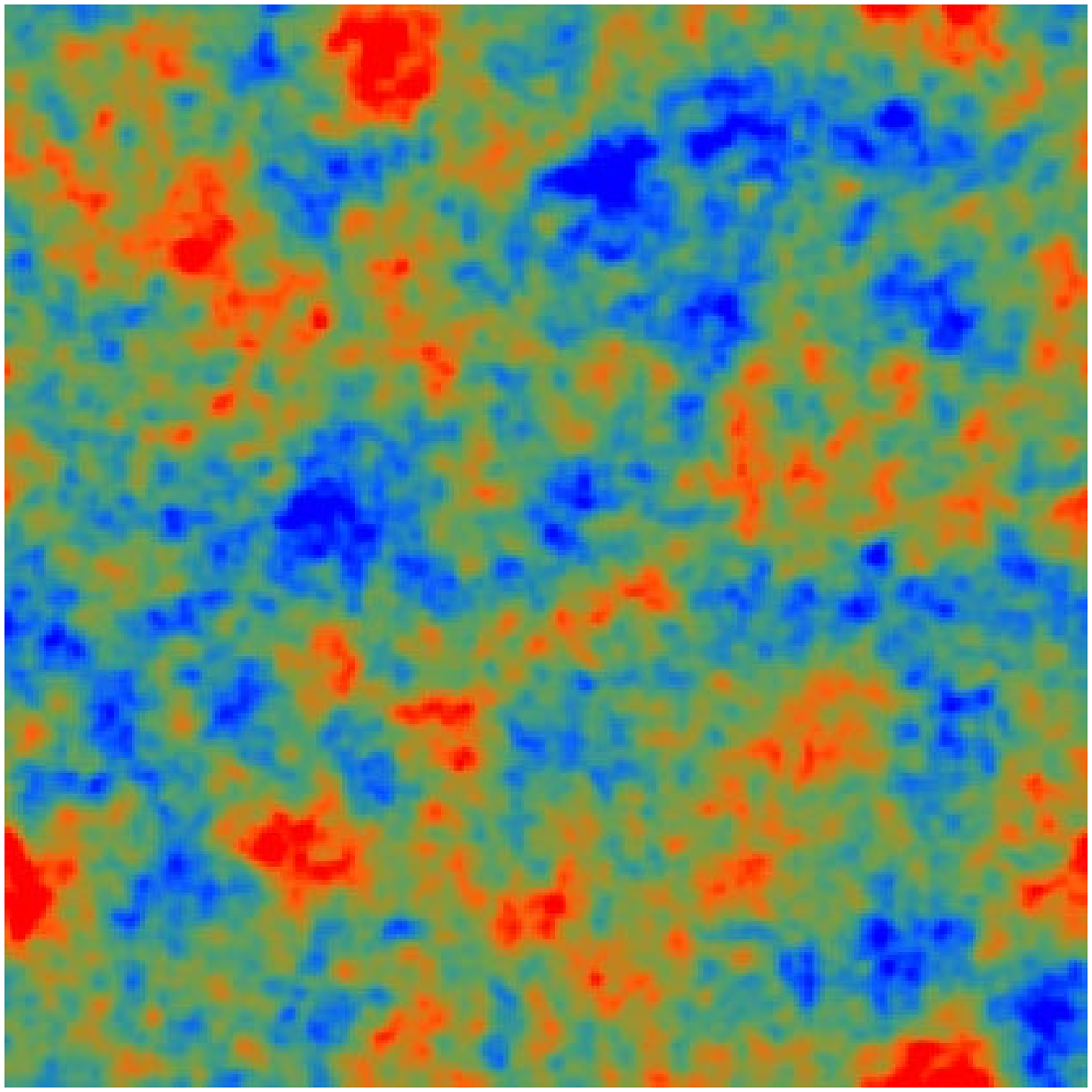}$\,$
\epsfxsize=2.25truein\epsffile{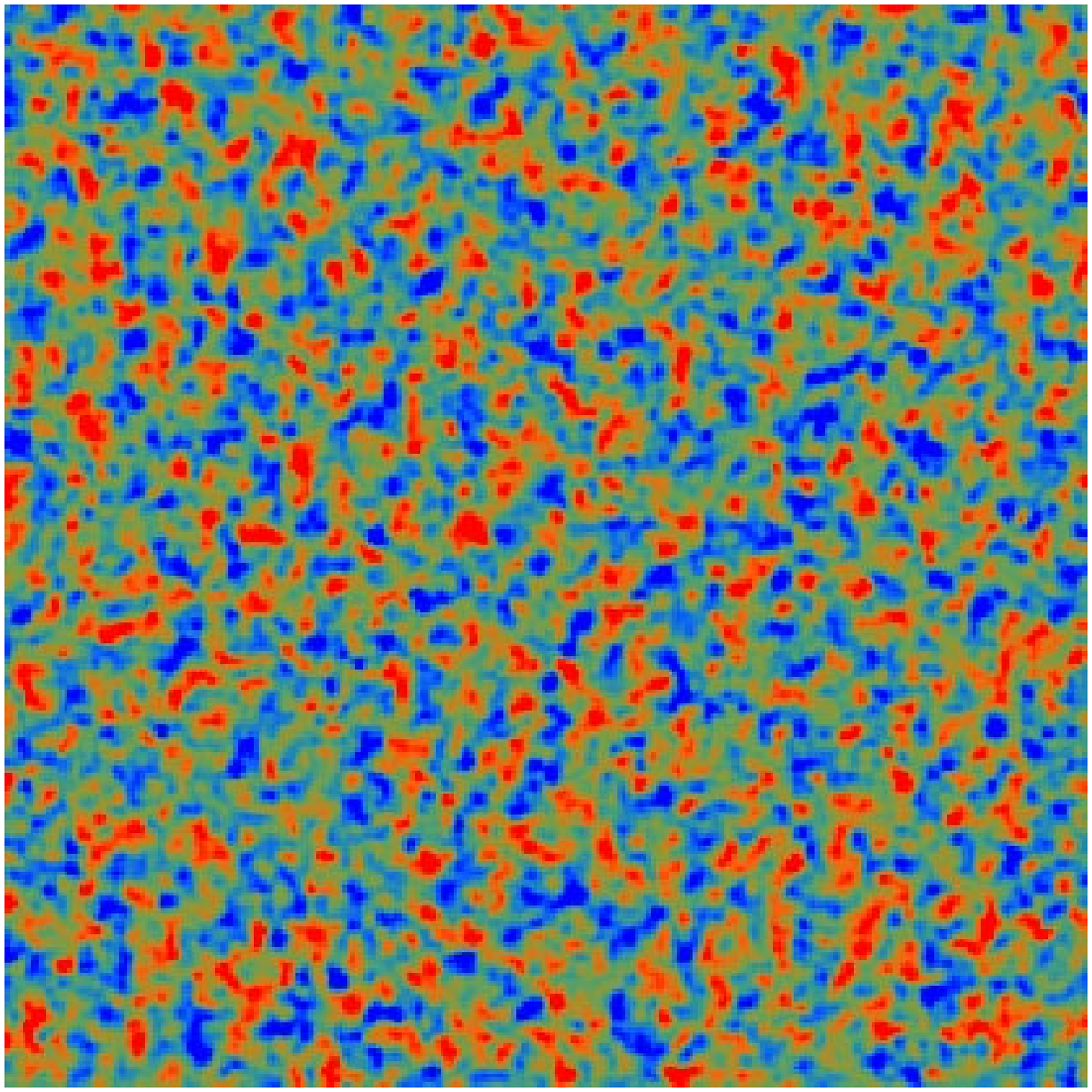}$\,$
\epsfxsize=2.25truein\epsffile{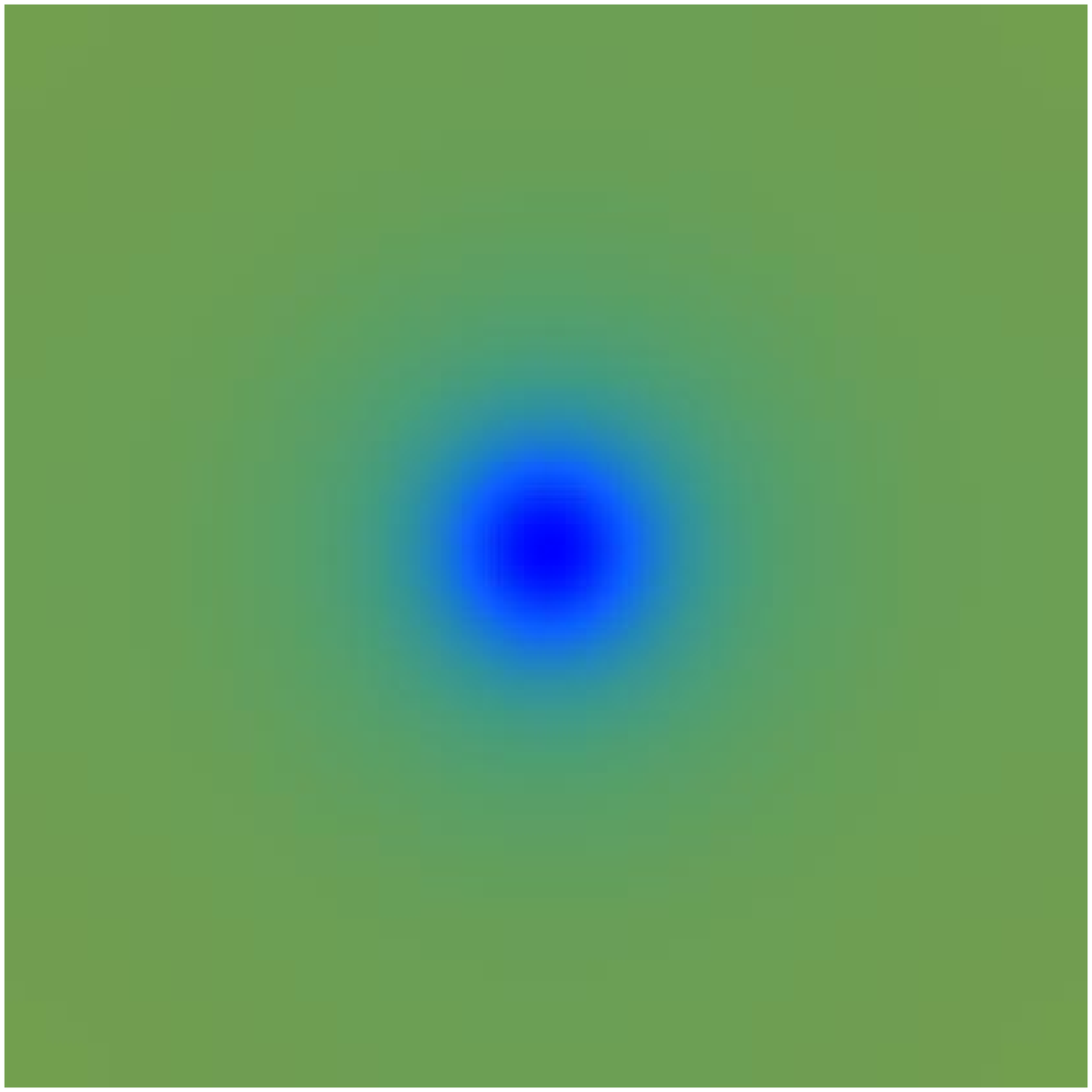}}
\vskip 0.1cm
\centerline{
\epsfxsize=2.25truein\epsffile{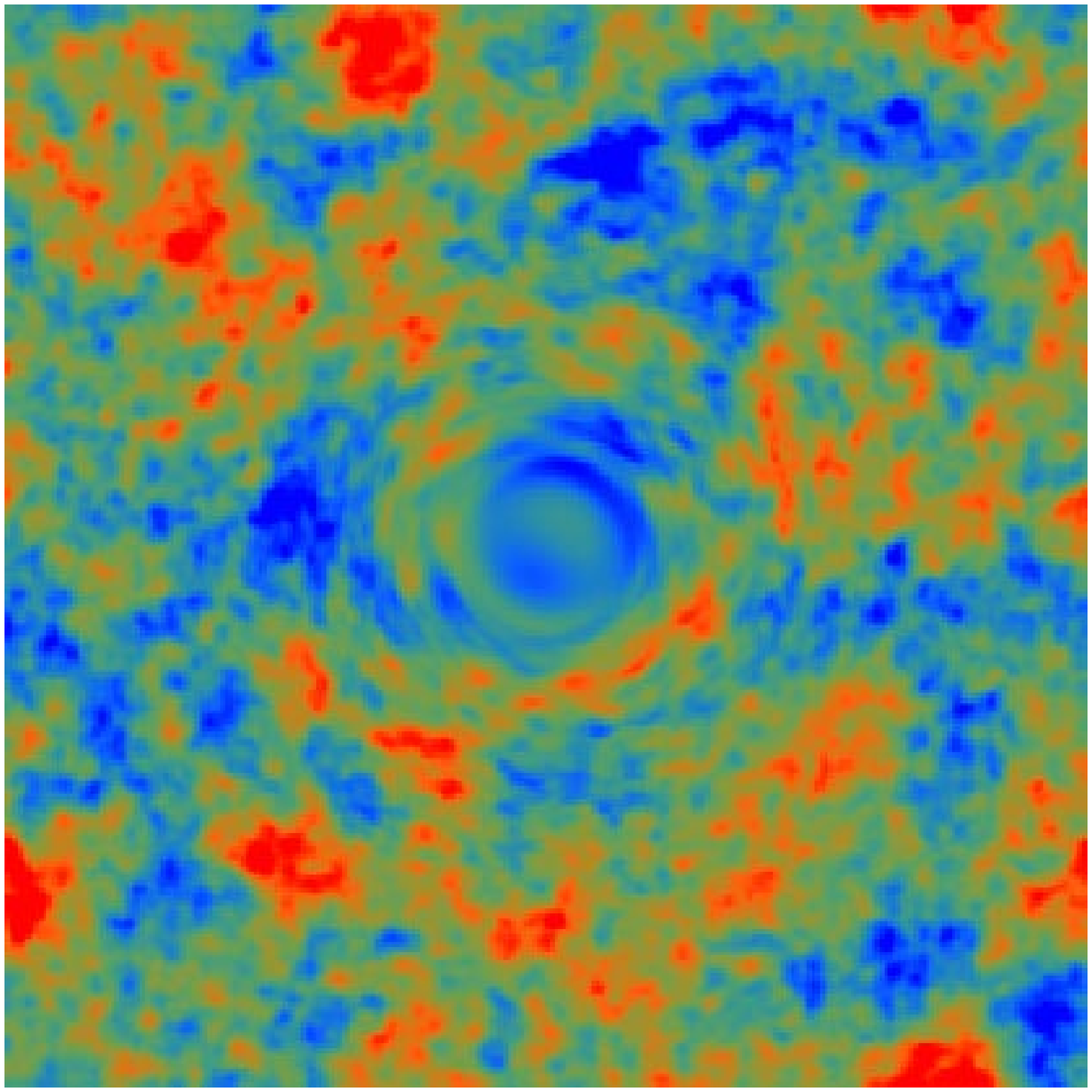}$\,$
\epsfxsize=2.25truein\epsffile{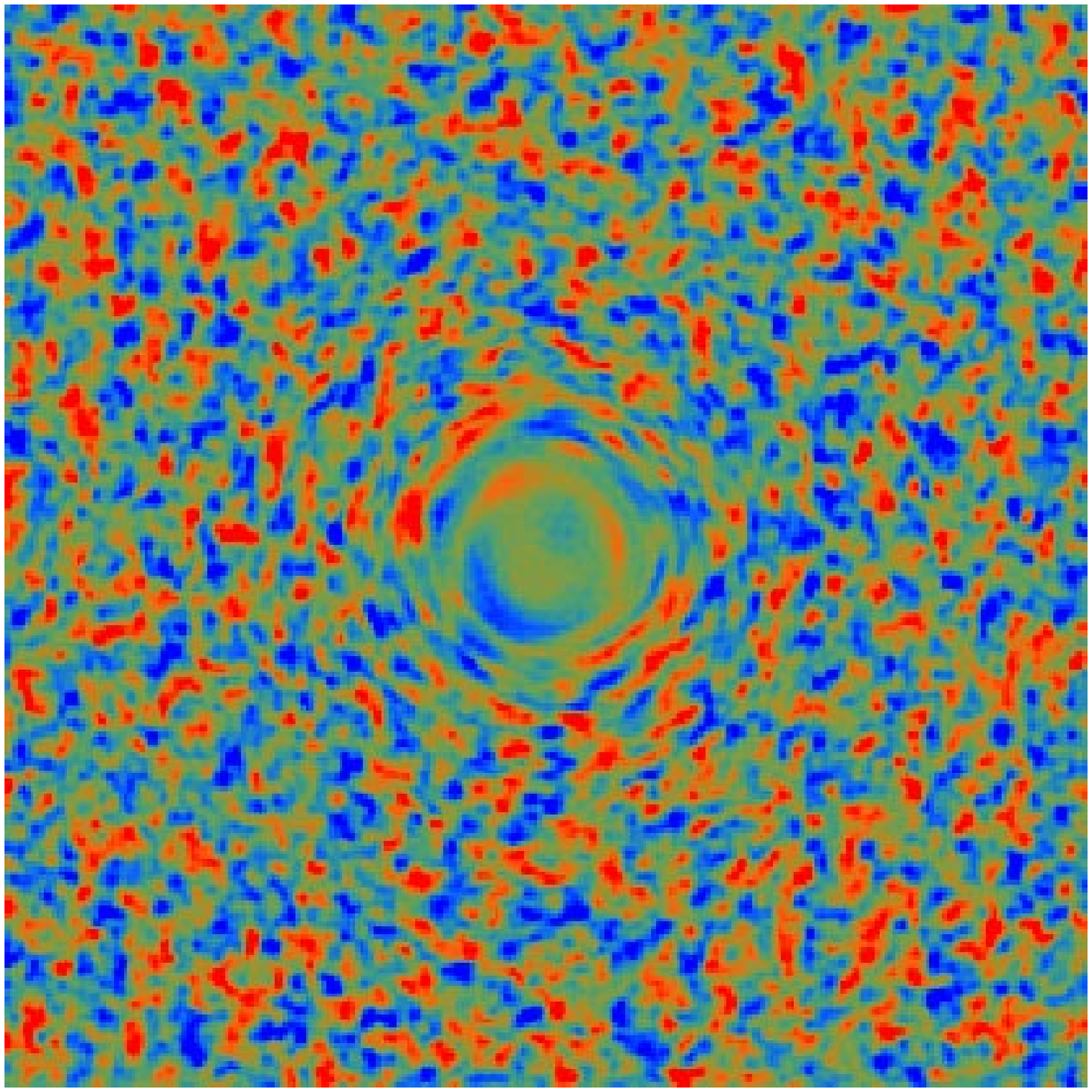}$\,$
\epsfxsize=2.25truein\epsffile{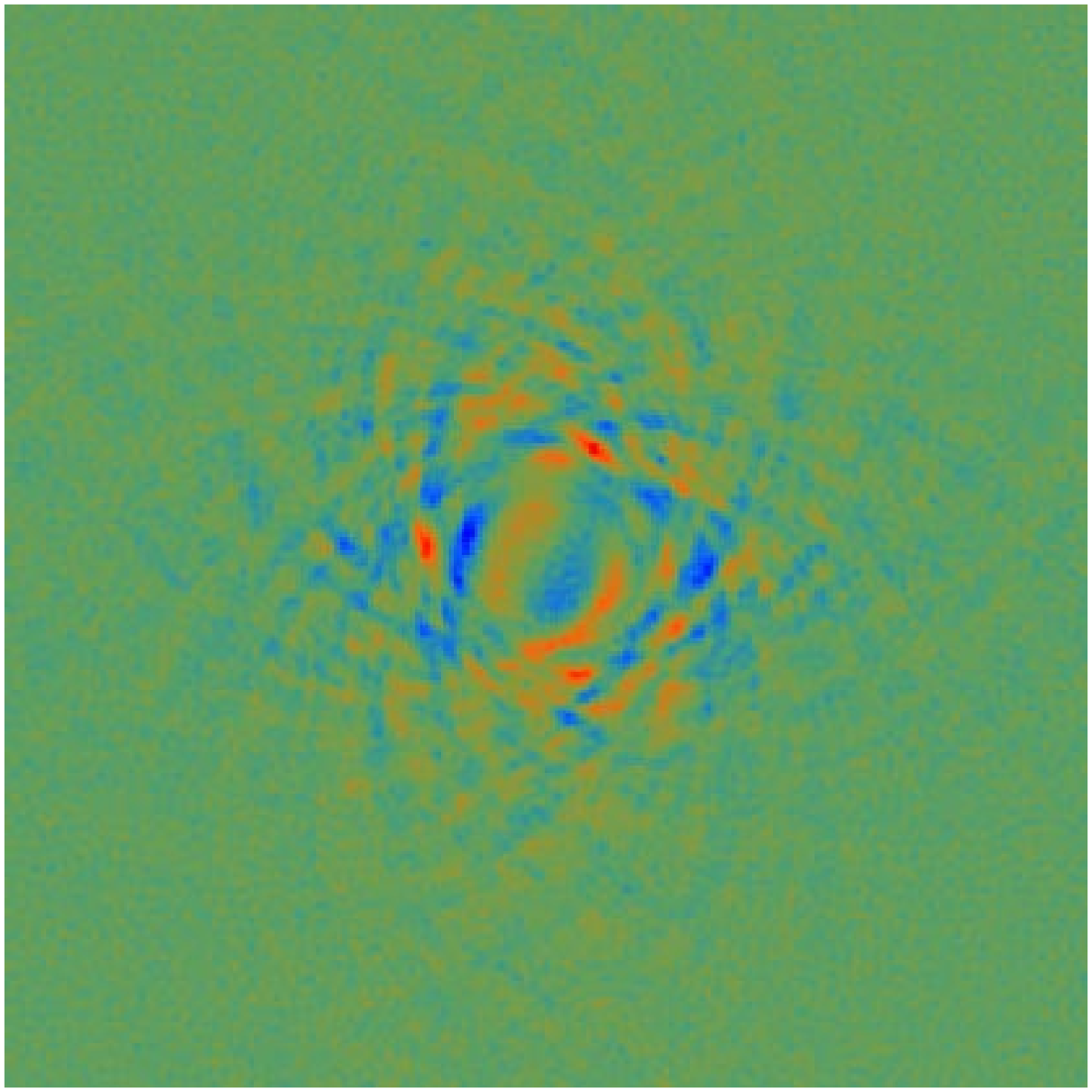}
}
\caption{An exaggerated example of the lensing effect on a $10^\circ \times
10^\circ$ field.  Top: (left-to-right)
unlensed temperature field, unlensed $E$-polarization field, spherically
symmetric deflection field $d(\bf n)$.      Bottom: (left-to-right)  
lensed temperature field, lensed $E$-polarization field, lensed $B$-polarization
field. The scale for the polarization and temperature fields differ by
a factor of 10.}
\label{fig:lensedmaps}
\end{figure*}

\section{Lensing}
\label{sec:lensing}
 
Weak lensing by the large-scale structure of the Universe remaps
the primary temperature field $\Theta(\bn) = \Delta T(\bn)/T$ and
dimensionless Stokes parameters $Q(\bn)$ and $U(\bn)$ as 
(\cite{BlaSch87} 1987; \cite{Ber97} 1997; \cite{ZalSel98} 1998)  
\begin{eqnarray}
\Theta(\bn) & = &  \tilde\Theta(\bn + {\bf d}(\bn)) \,,\\
\, [Q\pm iU](\bn) & = &  [\tilde Q\pm i \tilde U](\bn + {\bf d}(\bn))\,, \nonumber
\end{eqnarray}
where 
$\bn$ is the direction on the sky,
tildes denote
the unlensed field, and
${\bf d}(\bn)$ is the deflection angle.  It is related to the line of
sight projection of the gravitational potential $\Psi({\bf x},D)$ as
${\bf d} = \nabla \phi$,
\begin{equation}
\phi(\bn) = -2 \int d D\, {(D_s-D) \over D\, D_s} \Psi(D \bn,D)\,,
\end{equation}
where $D$ is the comoving distance along the line of sight in  the 
assumed flat cosmology and $D_s$ denotes the distance to the
last-scattering surface.  In the fiducial cosmology the rms deflection is
$2.6'$ but its coherence is several degrees.

We will work mainly in harmonic space and consider sufficiently small
sections of the sky such that spherical harmonic moments of order $(l,m)$ 
may be replaced by plane waves of wavevector $\bl$.  The all-sky 
generalization will be presented in a 
separate work (Okamoto \& Hu, in prep).
In this case, the temperature,
polarization, and potential fields may be decomposed as
\begin{eqnarray}
\Theta(\bn) & = & 
		\intl{}
		\Theta(\bl) e^{i \bl \cdot \bn} \,,\\
\label{eqn:multipoles}
\, [Q\pm i U](\bn) &=& 
		- 
		\intl{} [E(\bl)\pm i B(\bl)] e^{\pm 2i\varphi_{\bf l}} e^{i \bl \cdot \bn} \,,
		\nonumber\\
\phi(\bn) &= & 
		\int { d^2 L \over (2\pi)^2}
		\phi(\bll) e^{i \bll \cdot \bn} \,,\nonumber
\end{eqnarray}
where $\varphi_{\bl} = \cos^{-1} (\hat {\bf x} \cdot \hat{\bf l})$.
Lensing changes the Fourier moments by (\cite{Hu00b} 2000b)
\begin{eqnarray}
\delta \Theta(\bl) &=& \intlp{} \tilde\Theta(\bl') W(\bl',\bll) \,,\\
\label{eqn:lensedl}
\delta E(\bl)      &=& \intlp{} 
\Big[ \tilde E(\bl') \cos 2\varphi_{\bl'\bl}
     -\tilde B(\bl') \sin 2\varphi_{\bl'\bl} \Big]
W(\bl',\bll)\,,\nonumber\\
\delta B(\bl)      &=& \intlp{} 
\Big[ \tilde B(\bl') \cos 2\varphi_{\bl'\bl}
     +\tilde E(\bl') \sin 2\varphi_{\bl'\bl}\Big]
W(\bl',\bll)\,,\nonumber
\end{eqnarray}	
where $\varphi_{\bl'\bl} \equiv \varphi_{\bl'} - \varphi_{\bl}$,
$\bll = \bl - \bl'$, and
\begin{equation}
W(\bl,\bll) = -[{\bl \cdot \bll}]\phi(\bll)\,.
\end{equation}
Here $\delta \Theta = \Theta - \tilde \Theta$ for example.
In Fig.~\ref{fig:lensedmaps}, we show a toy example of the
effect of lensing on the temperature and polarization fields 
(see also \cite{BenBervan01} 2001). 
The effect on the $E$-polarization is similar to that of the temperature
and reflects the fact that $\cos 2\varphi_{\bl '\bl} \approx 1$
for $L \ll l$, where the lens is smooth compared with the field. 
Even in the absense of an unlensed $B$-polarization, lensing
will generate it.  The lensing structure differs since $\sin 2\varphi_{\bl'\bl}
\approx 0$ for $L \ll l$.  This fact will ultimately lead to a different 
range in $L$ of sensitivity to $\phi$ from the various fields.

\begin{figure}[t]
\centerline{\epsfxsize=3.5truein\epsffile{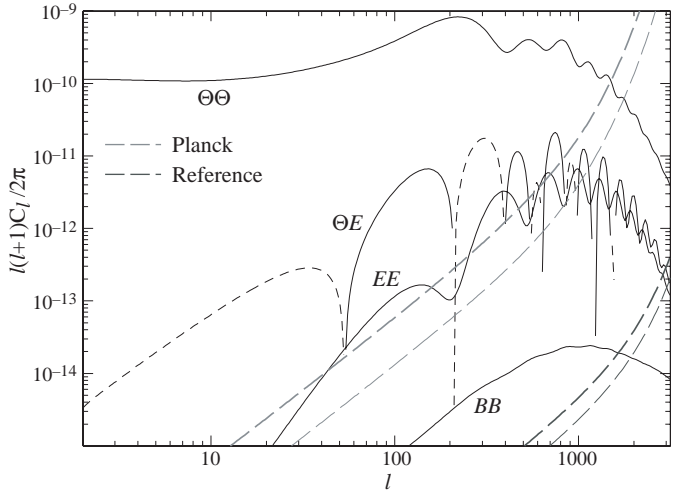}}
\label{fig:spectrum}
\caption{Power spectra of the CMB temperature and polarization fields
compared with the detector noise of the Planck satellite and a nearly
perfect experiment with a noise level of $\Delta_T = \Delta_P/\sqrt{2}=
1 \mu$K-arcmin and a beam of $\sigma=4'$ (long dashed lines, thick for polarization, thin for temperature).  The Planck experiment
has sufficient signal-to-noise to map the $\Theta$ field
but can only marginally map the $E$-polarization field;
the nearly perfect experiment can map the all three fields to $l = 
2000$.}
\end{figure}

Since the unlensed fields and potential perturbations
are assumed to be Gaussian and statistically isotropic, the 
statistical properties of the 
lensed fields may be completely defined by the unlensed power spectra
\begin{eqnarray}
\langle \tilde x^*(\bl) \tilde x(\bl') \rangle 
   &\equiv& (2\pi)^2 \delta(\bl-\bl') \tilde C_l^{xx'} \nonumber\,,\\
\langle \phi^*(\bll) \phi(\bll') \rangle 
   &\equiv& (2\pi)^2 \delta(\bll-\bll') L^{-2} C_L^{dd} \nonumber\,,
\end{eqnarray}
where $x = \Theta$, $E$, $B$ and we have chosen to express
the potential power spectrum with a weighting appropriate for
the deflection field $d(\bn)$.
Under the assumption of parity invariance 
\begin{equation}
\tilde C_l^{\Theta B} = \tilde C_l^{EB} = 0\,,
\end{equation}
and in the absence of gravitational waves and vorticity $\tilde C_l^{BB}=0$.
The peak in the logarithmic power spectrum $L^2 C_L^{dd} / 2\pi$ at $L \sim 30-40$ defines the 
degree-scale coherence of the deflection angles.

Finally, we define the power spectra of the observed temperature
and polarization fields as
\begin{equation}
\langle x^*(\bl) x(\bl') \rangle 
   \equiv (2\pi)^2 \delta(\bl-\bl') C_l^{xx'} \nonumber\,,
\label{eqn:unlensed}
\end{equation}
where the power spectra include all sources of variance to the
fields including detector noise and residual foreground contamination 
added in quadrature.
We will include Gaussian random detector noise of the form (\cite{Kno95} 1995)
\begin{eqnarray}
C_l^{\Theta\Theta}\Big|_{\rm noise} &=&
	\left({\Delta_T \over T_{\rm CMB}} \right)^2
	 e^{l(l+1)\sigma^2 /8\ln 2}\nonumber\,,\\
C_l^{EE}\Big|_{\rm noise} = 
C_l^{BB}\Big|_{\rm noise} &=& 
	\left({\Delta_P \over T_{\rm CMB}} \right)^2
	e^{l(l+1)\sigma^2 /8\ln 2}\,,
\end{eqnarray}
where $\Delta_{T,P}$ parameterizes white detector noise, here in units of
$\mu$K-radian, $T_{\rm CMB}=2.728 \times 10^6 \,\mu$K, and
$\sigma$ is the FWHM of the beam.   We will often assume $\Delta_P = 
\sqrt{2} \Delta_T$ as appropriate for fully-polarized detectors.
In Fig.~\ref{fig:spectrum}, we compare the signal and noise
contributions to the total power spectra for the Planck satellite
experiment\footnote{http://astro.estec.esa.nl/Planck}
 (minimum variance channel weighting 
from \cite{CooHu00a} 2000; $\Delta_T \approx 27\mu$K-arcmin,
$\Delta_P \approx 40\sqrt{2}\,\mu$K-arcmin, $\sigma\approx 7'$) and a
near perfect reference experiment ($\Delta_T = \Delta_P/\sqrt{2} = 1\mu$K-arcmin
and $\sigma=4'$).  
In general where the 
signal exceeds the noise power spectrum of a field, 
there is sufficient signal-to-noise
for mapping.  When this is not the case,
a statistical detection of the signal may still be possible.
The Planck experiment is on the threshold of being able to map the
$E$-polarization. The reference experiment can map all 3 fields
to $l \sim 2000$.

\section{Minimum Variance Estimators}
\label{sec:minvar}

As can be seen from Eqn.~(\ref{eqn:lensedl}), lensing mixes and therefore
correlates the Fourier modes across a range defined by the power in the
deflection field $C_L^{dd}$ (\cite{Hu00b} 2000b).  
Consider averaging over an ensemble of
realizations of the temperature and polarization
fields but with a fixed lensing field.  The two-point correlation
of the modes takes the form
\begin{equation}
\langle x(\bl) x'(\bl') \rangle_{\rm CMB} = f_\alpha(\bl,\bl') \phi({\bll})\,,
\label{eqn:cmbavg}
\end{equation}
where $x, x' = \Theta, E, B$, and $\bll =\bl+\bl'$.
We have assumed $\bl \ne -\bl'$ and will
use the subscript $\alpha$ to distinguish between
choices of the $xx'$ pairing, e.g. $\alpha = \Theta\Theta$.  
The correlation returns the value of
the deflection potential with weightings $f_\alpha$ that depend on the 
unlensed power spectra of
Eqn.~(\ref{eqn:unlensed}), which are given explicitly in Tab. 1.

\begin{table}
\begin{center}
\begin{tabular}{ll}
$\alpha$ 		& $f_\alpha({\bl_1,\bl_2})$ \vsp
\hline
$\Theta\Theta$	& $\tilde C_{l_1}^{\Theta\Theta}  \dotfac{1}
	        +  \tilde C_{l_2}^{\Theta\Theta}  \dotfac{2}$\vsp
$\Theta E$	& $\tilde C_{l_1}^{\Theta E}\cos 
		  \varphi_{\bl_1\bl_2}
		  \dotfac{1}
    	        +  \tilde C_{l_2}^{\Theta E}\dotfac{2}$\vsp
$\Theta B$	& $\tilde C_{l_1}^{\Theta E}\sin 2
			\varphi_{\bl_1\bl_2}
		  \dotfac{1}
		  $\vsp
$E E$	        & $[\tilde C_{l_1}^{E E}
		  \dotfac{1}
	          +\tilde C_{l_2}^{E E}
		  \dotfac{2}
			]\cos 2
			\varphi_{\bl_1\bl_2}
			$ \vsp
$E B$		& $[\tilde C_{l_1}^{E E}
		  \dotfac{1}
		 -\tilde C_{l_2}^{B B}
		  \dotfac{2}]
		\sin 2
			\varphi_{\bl_1\bl_2}
		$ 
		\vsp
$B B$	        & $[\tilde C_{l_1}^{B B}
		  \dotfac{1}
	         +\tilde C_{l_2}^{B B}
		  \dotfac{2}]
			\cos 2
			\varphi_{\bl_1\bl_2}
		$ \vsp
\end{tabular}
\end{center}
\caption{Minimum variance filters}
\end{table}

The two point correlations of the CMB Fourier modes themselves 
cannot be used to reconstruct the deflection potential since
$\phi$ is also statistically isotropic so that in the true ensemble average
$\langle \phi(L) \rangle = 0$.
Eqn.~(\ref{eqn:cmbavg}) does suggest however that an appropriate
average over pairs of multipole moments can be used to estimate
the deflection field $d({\bn})$.

Let us define a general weighting of the moments
\begin{eqnarray}
d_\alpha({\bll}) =  {A_\alpha(L) \over L} \intl{1} x(\bl_1) x'(\bl_2)  
F_\alpha(\bl_1,\bl_2)\,,
\label{eqn:estimator}
\end{eqnarray}
where $\bl_2 = \bll -\bl_1$ and the normalization 
\begin{eqnarray}
A_\alpha(L) =  L^2 \Bigg[ \intl{1} f_\alpha(\bl_1,\bl_2)
F_\alpha(\bl_1,\bl_2) \Bigg]^{-1} \,.
\end{eqnarray}
is chosen so that
\begin{equation}
\langle d_\alpha(\bll) \rangle_{\rm CMB} = d({\bll}) \equiv L \phi({\bll})\,.
\end{equation}
In general there are 6 estimators corresponding to the
$3!$ pairs of $\Theta$, $E$, $B$.  In the assumed cosmology, where
gravitational wave perturbations are negligible compared with
density perturbations, $\alpha=BB$ has vanishing signal-to-noise effectively
reducing the estimators to 5.

We can optimize the filter $F_\alpha$ by minimizing the variance
$\langle d_\alpha^*(L) d_\alpha(L) \rangle$, subject to the
normalization constraint
\begin{equation}
F_\alpha(\bl_1,\bl_2)= 
{
C_{l_1}^{x'x'} C_{l_2}^{xx} f_\alpha(\bl_1,\bl_2)
-
C_{l_1}^{x x'} C_{l_2}^{xx'} f_\alpha(\bl_2,\bl_1)
\over
C_{l_1}^{xx} C_{l_2}^{x'x'} C_{l_1}^{x'x'} C_{l_2}^{xx}
- 
(C_{l_1}^{xx'} C_{l_2}^{xx'})^2} \,.
\label{eqn:optimal}
\end{equation}
This filter takes on simple forms for
two common cases: if $x=x'$, as in the case
of $\alpha=\Theta\Theta$, $E E$ and $B B$,
\begin{equation}
F_\alpha(\bl_1,\bl_2) \rightarrow {f_\alpha(\bl_1,\bl_2) \over
	 2 C_{l_1}^{xx} C_{l_2}^{xx}};
\end{equation}
if $\tilde C_l^{xx'}=0$, as in the case of $\alpha=\Theta B$ and $E B$,
\begin{equation}
F_\alpha(\bl_1,\bl_2) \rightarrow {f_\alpha(\bl_1,\bl_2) \over 
	C_{l_1}^{xx} C_{l_2}^{x'x'}}.
\end{equation}

\begin{figure}[t]
\centerline{
\epsfxsize=3.5truein\epsffile{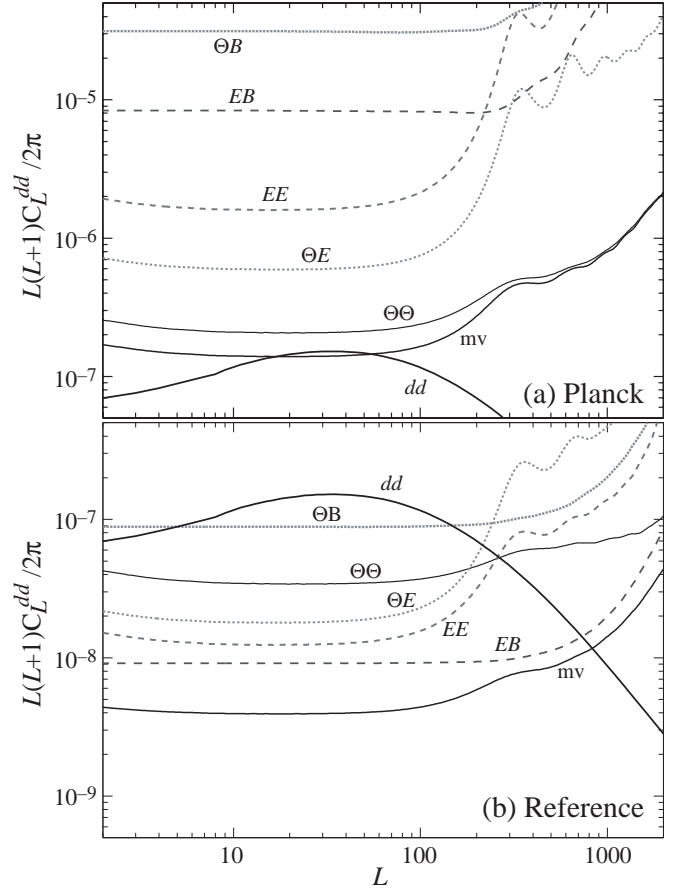}}
\caption{Deflection signal ($dd$) and noise power spectra of the 
quadratic estimators and their minimum variance (mv) combination:
(a) Planck experiment  (b) reference experiment.
As the sensitivity of the experiment improves the best
quadratic estimator switches from $\Theta\Theta$ to 
$EB$.  
Only the $EB$-estimator can reconstruct the
mass distribution at $L \gsim 200$.
}
\label{fig:noise1}
\end{figure}

The noise properties of these estimators follows from 
\begin{eqnarray}
\langle d_\alpha^*({\bll}) d_\beta({\bll}') \rangle
&=& (2\pi)^2 \delta({\bll}-{\bll'}) 
[C_L^{dd} + N_{\alpha\beta}(L)]\,,
\end{eqnarray}
where
\begin{eqnarray}
N_{\alpha\beta}(L) &=& L^{-2} A_\alpha(L) A_\beta(L)  
\intl{1} 
		F_\alpha (\bl_1,\bl_2)
	\Big( 
		F_\beta (\bl_1,\bl_2)
\nonumber\\
&&\times 
C_{l_1}^{x_\alpha x_\beta}  C_{l_2}^{x_\alpha' x_\beta'} + 
		F_\beta (\bl_2,\bl_1)
                  C_{l_1}^{x_\alpha x_\beta'} C_{l_2}^{x_\alpha' x_\beta} \Big) \,.
\label{eqn:covariance}
\end{eqnarray}
Recall that the $xx$-power spectra account for both the cosmic
variance of the fields and the noise variance of the experiment.
Notice that for the minimum variance filter
\begin{equation}
N_{\alpha\alpha}(L) = A_{\alpha}(L)\,.
\end{equation}
In Fig.~\ref{fig:noise1}, we compare the signal and noise power
spectra for the Planck experiment and the reference experiment 
defined in \S \ref{sec:lensing}. 
Recall that true mapping is possible when the
signal exceeds the noise spectrum.
For the Planck experiment, $\Theta\Theta$ provides the best
estimator reflecting the fact that Planck will not be able to 
produce true maps of the polarization modes.  
Furthermore, the signal-to-noise is highest at $L \lsim 200$
reflecting the fact the modes are mainly correlated across 
$\Delta L \sim 60$, where the deflection power spectrum peaks.

For the reference experiment, all 5 estimators have sufficient
signal-to-noise to produce maps at $L \lsim 200$.  
The $EB$ estimator has the
best signal-to-noise, and allows for mapping to $L \lsim 1000$.
The reason is that there is no noise variance contributed by an
unlensed $B$ field.  Furthermore, the signal intrinsically
comes from higher $L$.  A $B$-field at
a wavenumber $\bl$ cannot be generated from neighboring modes
$\bl' \sim \bl$ from the low $L$ deflection field
because of the $\sin$ term in the lensing kernel
(see Eqn.~\ref{eqn:lensedl}).  Thus the signal to noise is
relatively higher at high $L$ in the $EB$ estimator. 

\begin{figure}[t]
\centerline{
\epsfxsize=3.5truein\epsffile{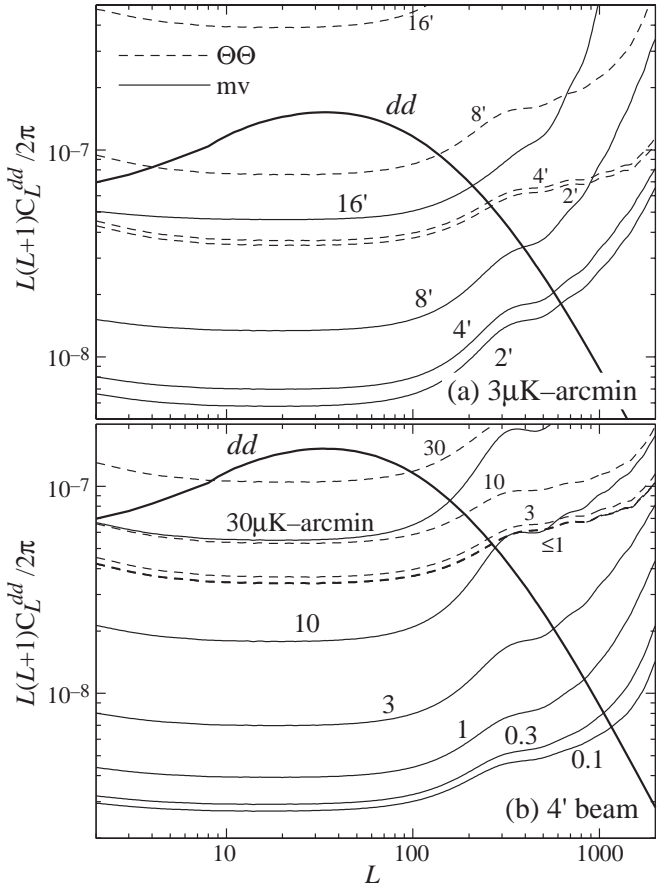}}
\caption{Deflection signal ($dd$) and noise power spectra for the minimum
variance (mv; solid lines) and $\Theta\Theta$ (dashed lines) estimators
as a function of (a) beam size
$\sigma$ and (b) noise level $\Delta_T=\Delta_P/\sqrt{2}$.  The 
noise saturates to its minimum by $\sigma \approx 2-4'$
and $\Delta_T \approx 0.1-0.3 \mu$K-arcmin as
the polarization field is mapped to the cosmic variance limit
out to $l < 2000$.} 
\label{fig:noise2}
\end{figure}

\begin{figure*}[t]
\centerline{
\epsfxsize=2.25truein\epsffile{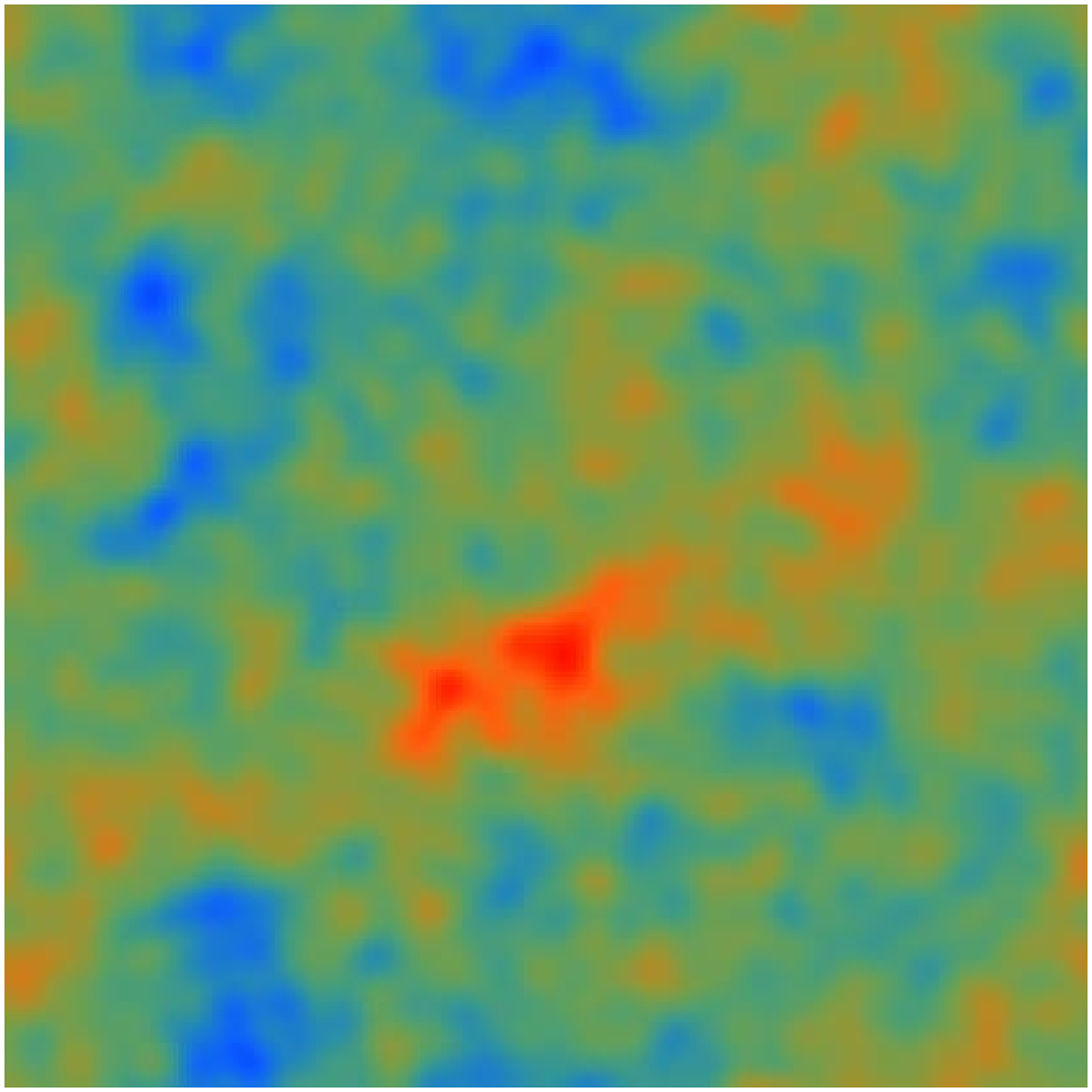}$\,$
\epsfxsize=2.25truein\epsffile{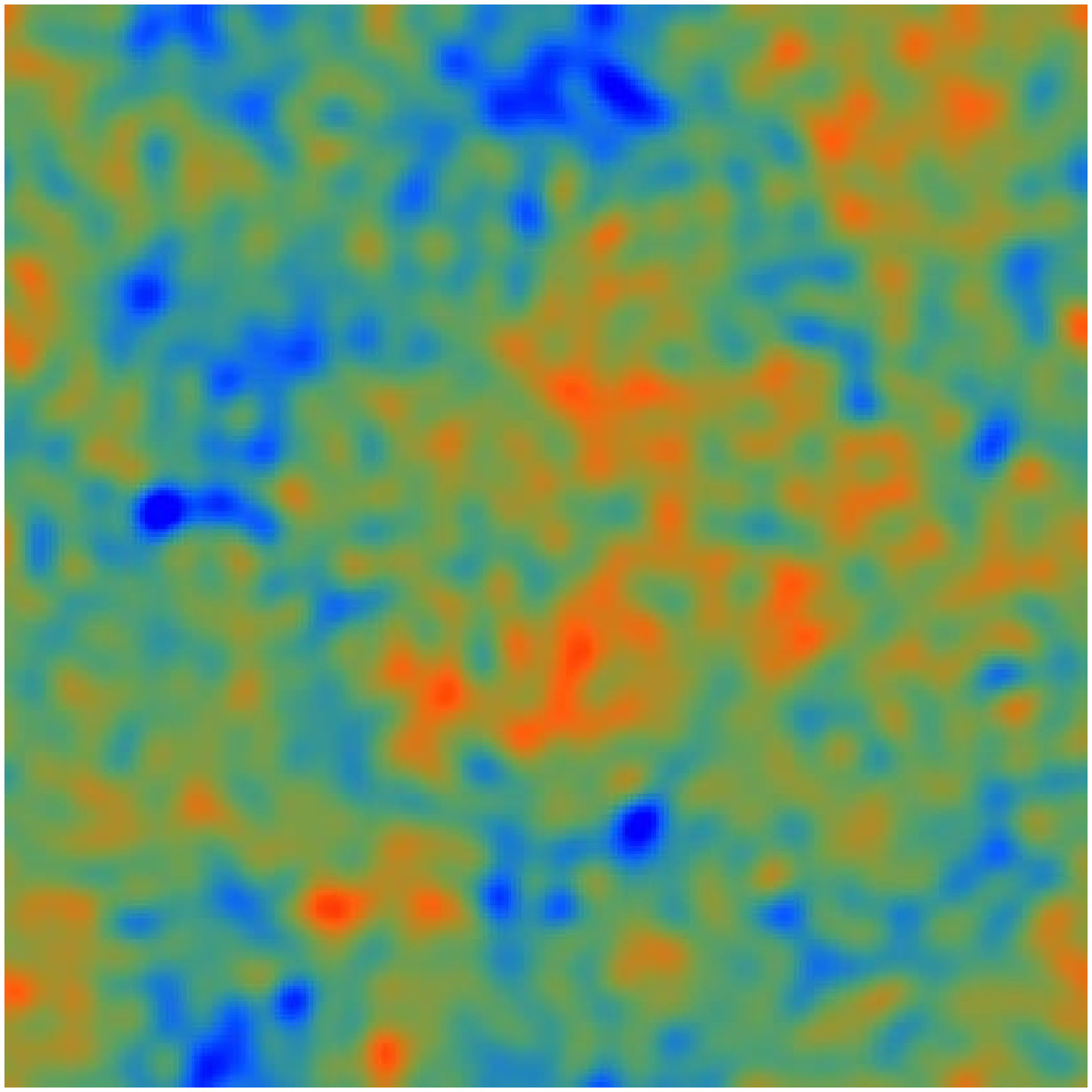}$\,$
\epsfxsize=2.25truein\epsffile{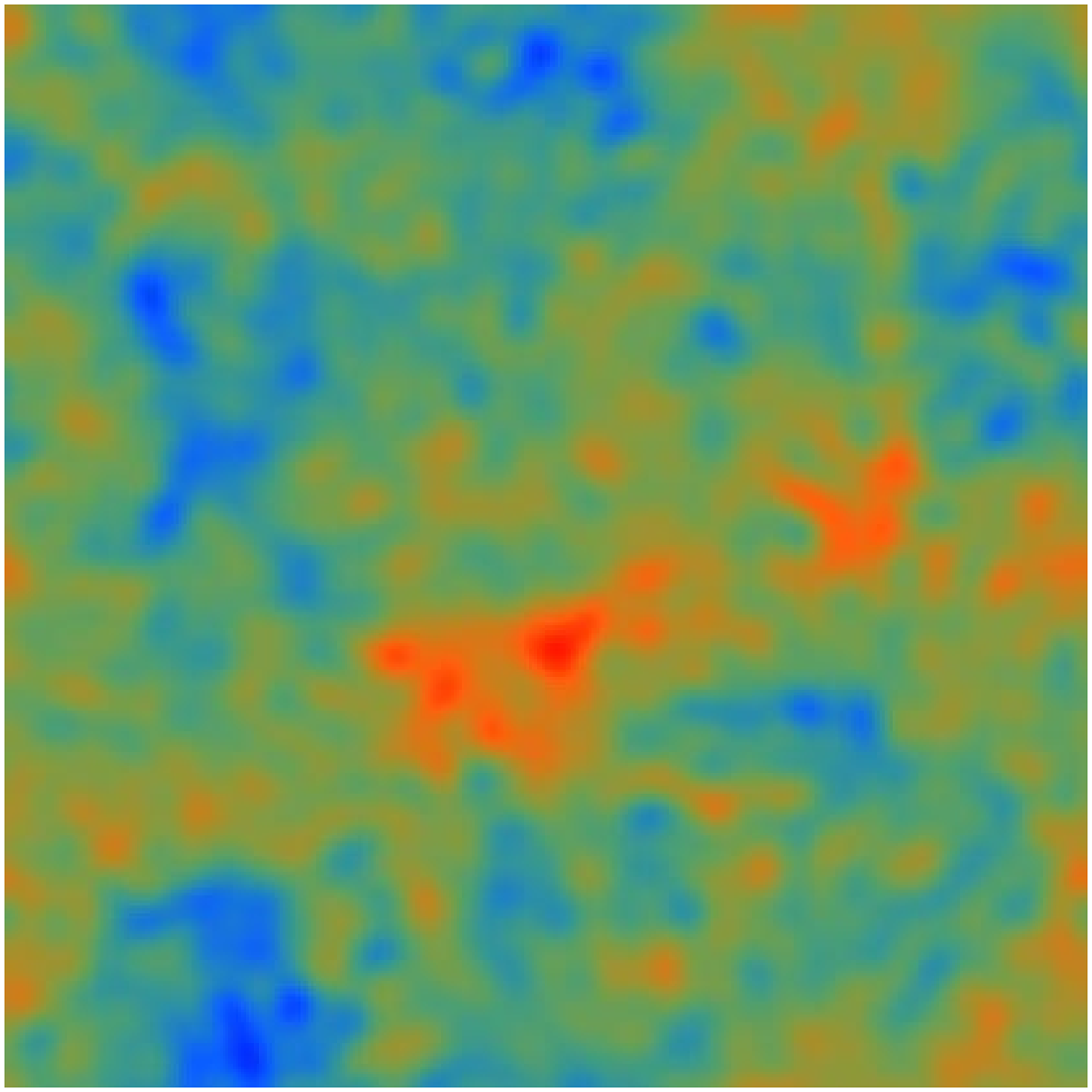}}
\caption{Mass reconstruction on a $10^\circ \times 10^\circ$ field
with the reference experiment 
($\Delta_T=\Delta_P/\sqrt{2}=1 \mu$K-arcmin and $\sigma=4'$):
(a) deflection field, (b) $\Theta\Theta$-reconstruction, 
(c) $EB$-reconstruction.}
\label{fig:reconmaps}
\end{figure*}

For experiments that are intermediate in sensitivity between
Planck and the reference experiment, the five estimators of the
deflection field have comparable signal-to-noise and may
be used to cross check each other.  At high-$L$ where the
individual estimators are noise limited, combining
the estimators as
\begin{equation}
d_{\rm mv}({\bll}) = \sum_\alpha w_\alpha(L) d_\alpha({\bll})\,,
\end{equation}
can substantially reduce the noise.  The minimum
variance weighting is a generalization of the inverse variance
weighting that accounts for the covariance 
in Eqn.~(\ref{eqn:covariance})
\begin{equation}
w_\alpha = {\sum_\beta ({\bf N}^{-1})_{\alpha\beta} 
	\over \sum_{\beta\gamma} ({\bf N}^{-1})_{\beta\gamma}}\,.
\end{equation}
The noise variance
\begin{equation}
\langle d_{\rm mv}^*(\bll) d_{\rm mv}(\bll') \rangle
= (2\pi)^2 \delta(\bll-\bll') 
[C_L^{dd} + N_{\rm mv}(L)]
\end{equation}
becomes
\begin{equation}
N_{\rm mv} = {1 \over \sum_{\alpha\beta} ({\bf N}^{-1})_{\alpha\beta}}\,.
\end{equation} 
Note that the $\Theta\Theta$ and $EB$ estimators are independent and
those estimators that are correlated have correlation coefficients 
$N_{\alpha\beta}/\sqrt{N_{\alpha\alpha} N_{\beta\beta}}$ of no more than
tens of percents.

The minimum variance noise spectra for Planck and the reference experiment 
are shown in Fig.~\ref{fig:noise1}.
We give it as a function of the noise $\Delta_T$ and beam
$\sigma$ in Fig.~\ref{fig:noise2}.  The signal-to-noise
saturates around $\Delta_T \approx 0.1-0.3 \mu$K-arcmin 
and $\sigma \approx 2-4'$.  
Below $\Delta_T \approx 26\mu$K-arcmin
($\Delta_P \approx 37\mu$K-arcmin), the combined
signal-to-noise in the $\Theta E$ and $E E$ estimators
exceeds that in $\Theta\Theta$ at $L \approx 40$ and $\sigma=4'$.
For the smaller scales of $L \approx 300$, where only the $EB$ estimator
plays a role, the $EB$ and $\Theta\Theta$ estimators
have comparable signal-to-noise around 
$\Delta_T \approx 6\mu$K-arcmin
($\Delta_P \approx 8\mu$K-arcmin)
at $L \approx 300$ and $\sigma=4'$.

\section{EB Estimator}
\label{sec:eb}

As we have seen in the previous section, the $EB$-estimator of
the deflection field has the potential to map the mass
distribution out to $L \approx 1000$.   We therefore explicitly construct
and test this estimator in this section.  This construction is
very similar to that of the $\Theta\Theta$ estimator presented
in \cite{Hu01b} (2001b).

From Eqn.~(\ref{eqn:estimator}), the $EB$ estimator is
\begin{equation}
d_{EB}({\bll}) = {A_{EB}(L) \over L} 
                  \intl{}  
		E(\bl) B(\bl')  
		  {
		\tilde C_{l}^{EE} 
		{\bll} \cdot \bl
		\over C_{l}^{EE} C_{l'}^{BB}}
		\sin 2\varphi_{\bl\bl'}\,,
\end{equation}
where recall ${\bf L} = \bl + \bl'$.  The convoluted form of this estimator
suggests that it may be re-expressed as a product of fields
on the sky.  To see this, rewrite
\begin{equation}
\sin 2\varphi_{\bl\bl'} = 2 ({\hat\bl \cdot \hat\bl'})[\bn \cdot (
				{\hat\bl \times \hat\bl'})]\,,
\end{equation}
where $\bn = -{\bf e}_3$. We can then define
the filtered fields
\begin{eqnarray}
E_{ijk}(\bn) &=& \intl{} 
	l (\hat l_i \hat l_j \hat l_k) {\tilde C_l^{EE} \over C_l^{EE}} E(\bl)  e^{i \bl \cdot \bn}\,,\\
B_{ij}(\bn)  &=& \intl{}
	  (\hat l_i \hat l_j) {1 \over C_l^{BB}} B(\bl)
  e^{i \bl \cdot \bn}\,.
\end{eqnarray}
There are $4$ unique filtered $E$-fields and $3$ unique filtered $B$-fields.
They may be combined to form the appropriate dot and cross products
\begin{equation}
G_i(\bn) = 2 \sum_{jkm} E_{ijk}(\bn) B_{jm}(\bn) \epsilon_{km3}\,,
\end{equation}
where $\epsilon_{ijk}$ is the Levi-Civita symbol.  
The deflection field is then reconstructed as 
\begin{equation}
d_{EB}(\bll) = -{A_{EB}(L) \over L} {\bll} \cdot {\bf G}({\bll})\,.
\end{equation}
The other quadratic estimators can be constructed in a similar
fashion.  

Fig. \ref{fig:reconmaps} shows an example of the $EB$ reconstruction 
compared with the $\Theta\Theta$ reconstruction on
a $10^\circ \times 10^\circ$ field with the reference experiment.
Notice that the $EB$-reconstruction has substantially lower
noise on small angular scales.  We assume here that the unlensed
power spectra have been determined externally from precision 
satellite missions and through the modelling with cosmological parameters
(see \cite{Hu01b} 2001b).  Errors in the determination translate
into non-optimal filters and a small bias in the amplitude of the
reconstructed maps.

\section{Applications}
\label{sec:applications}

\begin{figure}[t]
\centerline{\epsfxsize=3.5truein\epsffile{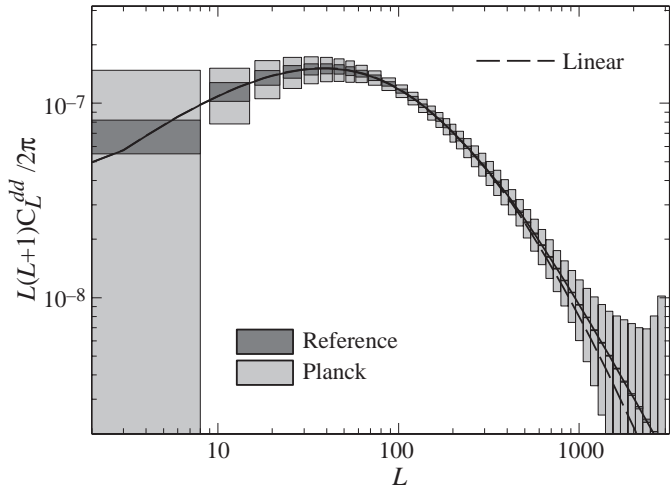}}
\caption{Statistical errors achievable on the deflection power spectrum
with the Planck ($f_{\rm sky}=0.65$) and reference experiments ($f_{\rm sky}=1$).  Boxes represent band averaging width and 1$\sigma$ errors.
The polarization information 
in the reference experiment allows for a cosmic variance limited measurement
of the projected power spectrum out to $L \sim 1000$.  In this regime,
the fluctuations are almost completely linear (dashed lines).}
\label{fig:derrors}
\end{figure}

In this section, we outline four applications for mass reconstruction:
measurement of the (linear) power spectrum in projection, 
cross correlation with cosmic shear observations, cross correlation 
with the temperature field, and decontamination of the polarization signature
of gravitational waves.  The first three applications have been
extensively discussed in \cite{Hu01c} (2001c) for the $\Theta\Theta$
temperature based estimator
and we refer the reader to details therein.   Here we focus on the
additional information provided by the polarization field.

\begin{figure}[t]
\centerline{\epsfxsize=3.5truein\epsffile{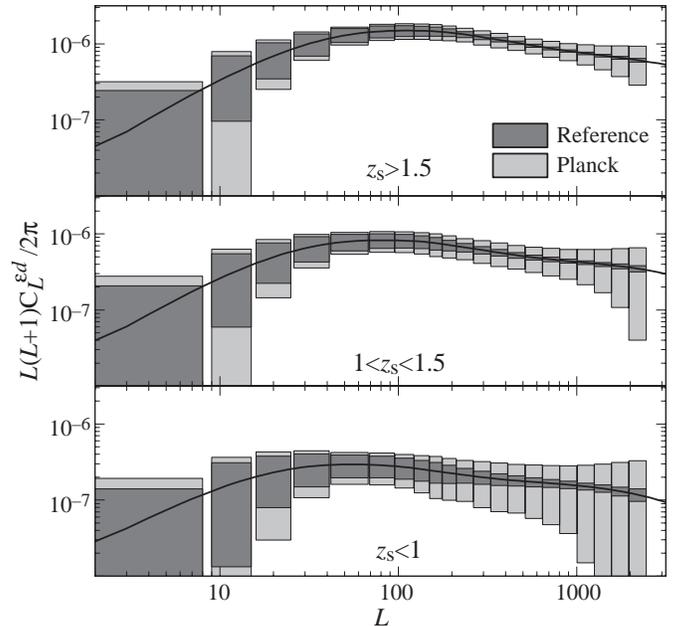}}
\caption{Statistical errors on the cross correlation of CMB deflections
and cosmic shear in three source redshift bands on a $1000$ deg$^{2}$ patch
of sky for the Planck and reference experiment.  
Assumptions for the cosmic shear experiment are given in the text.
Precision measurements from the polarization estimators enables highly
significant cross correlation detection and hence tomographic studies
of structure evolution.}
\label{fig:shear}
\end{figure}

\subsection{Linear Power Spectrum}

The most direct application of mass reconstruction is to measure
the matter power spectrum in projection, i.e. the deflection power
spectrum $C_L^{dd}$ itself.  Power spectrum measurement requires
only a statistical detection of the deflection field, not a true
reconstructed map and therefore can be extended to higher wavenumbers or 
smaller scales than is possible for mapping.  The
noise level for the estimation of band powers is reduced by 
averaging over ${\bf L}$ directions in a band $\Delta L$
\begin{equation}
\Delta C_L^{dd} \approx {1 \over \sqrt{ L\Delta L f_{\rm sky}}}
		[C_L^{dd} + N_{\rm mv}(L)]\,,
\end{equation}
where $f_{\rm sky}$ is the fraction of the sky covered by the experiment.
In this approximation, the noise is assumed to be Gaussian.  This
should be a good approximation where the sample variance of the lenses
dominates the noise variance.  Formally, the noise will be increasingly 
non-Gaussian at high $L$
as the estimator is constructed out of fewer arcminute scale 
temperature and polarization fluctuations.  Quantification of
this effect for the temperature based reconstruction show that its
effects are minor (\cite{Hu01} 2001a); a full treatment requires the 
consideration of the temperature-polarization
trispectrum (Okamoto \& Hu, in prep.).

Polarization enables two advances over
what can be achieved by the temperature field alone.  
As in the case of mapping, polarization enables precision 
measurements at small scales through the $EB$-estimator.
In Fig.~\ref{fig:derrors}, we compare the Planck experiment 
(with $f_{\rm sky}=0.65$) and the reference experiment (with $f_{\rm sky}=1$);
as seen in Fig.~\ref{fig:noise1}, 
former relies mainly on the $\Theta\Theta$-estimator and the
later on the $EB$-estimator.  The noise in the Gaussian approximation
approaches the sample variance
limit of $\Delta C_L^{dd}/C_L^{dd} = (L\Delta L f_{\rm sky})^{-1/2}$ on
the scales $L \lsim 1000$, i.e. a total of 1\% precision in each 1\% of
sky. 
This corresponds to scales in the matter power spectrum of 
$0.002 \lsim k \lsim 0.2$ in $h$/Mpc representing the whole
linear regime today.

Equally importantly, the polarization allows for sharp consistency
tests on the power spectrum measurements at $L \lsim 500$.  In
the reference experiment, all 5 estimators have sufficient signal-to-noise
to measure the power spectrum here.
It is highly unlikely that any unknown contaminant from foregrounds
or instrumental systematics would affect specific quadratic combinations of
the temperature, $E$-polarization, $B$-polarization, in the same way. 

\begin{figure}[t]
\centerline{\epsfxsize=3.5truein\epsffile{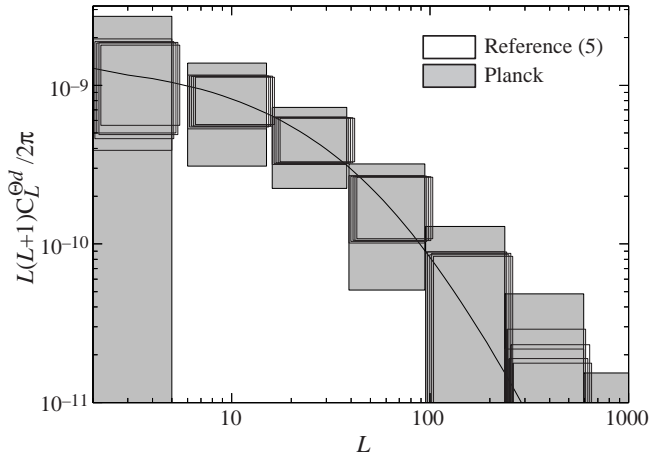}}
\caption{Statistical errors on the cross correlation of CMB deflections and
the temperature field for the Planck ($f_{\rm sky}=0.65$)  and reference 
($f_{\rm sky}=1$) experiments.  The five estimators of the deflection field
obtainable with polarization information enables five nearly independent,
cosmic variance limited detections of the cross correlation for $L < 100$ 
(shown offset slightly for clarity).
The cross correlation is extremely sensitive to the properties of the
dark energy.}
\label{fig:isw}
\end{figure}

\subsection{Evolution of Structure and Cosmic Shear}
\label{sec:shear}

One would like to go beyond the projected power spectrum to the
three-dimensional distribution to track the evolution of structure
and hence the physical properties of the dark matter and energy.
This is not possible through CMB lensing alone since the source
plane lies at the effectively infinite redshift of last-scattering.
Weak lensing also distorts the shape of distant, but for our
purposes foreground, galaxies allowing a measurement of the 
gradient of the deflection angles, or more properly, the 
so-called cosmic shear, from wide-field imaging surveys 
(see \cite{Mel99} 1999; \cite{BarSch01} 2001 for reviews).  
Since these sources are distributed across a range of redshifts,
the change in the mass reconstruction as a function of source
redshift can probe the radial distribution of matter 
tomographically (\cite{Witetal01} 2001). 
The effect on the power spectra and cross-correlation 
of cosmic shear $C_L^{\epsilon\epsilon}$
has been shown to be an effective probe of the dark energy
equation of state (\cite{Hut01} 2001; \cite{Hu01c} 2001c).
Because cosmic shear studies are most effective for $L \gsim 100$,
tomography with the lensing of the CMB temperature is difficult.

By extending the measurements to overlapping 
wavenumbers, CMB polarization allows 
tomographic studies to be anchored at the high-$z$ end.  
In Fig.~\ref{fig:shear}, we show the errors on the CMB 
deflection-cosmic shear cross power spectrum  $C_L^{\epsilon d}$ achievable
with the Planck vs. reference experiment and 
$1000$ deg$^2$ of overlap with a cosmic shear survey out to 
median source redshift $z=1$, divided into three redshift bands 
$z_s<1$, $1<z_s<1.5$, $z_s>1.5$.  Errors on the cosmic shear side assume 
$n = 56$ gal/arcmin$^2$, and an intrinsic shear measurement error of 
$\langle \gamma_{\rm int} \rangle^2 = 0.4$ per component per galaxy.

\begin{figure}[t]
\centerline{
\epsfxsize=2.5truein\epsffile{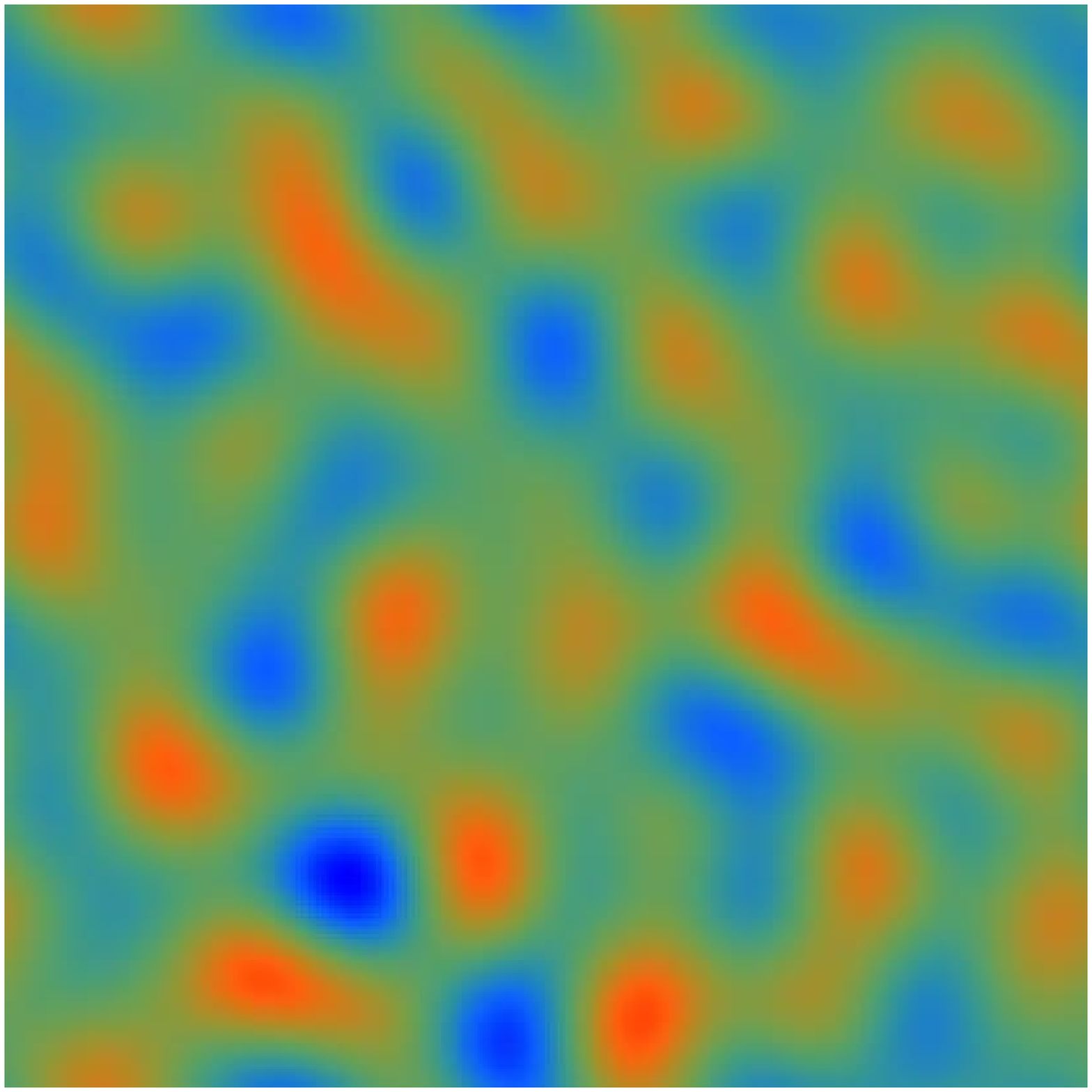}}
\vskip 0.1cm
\centerline{
\epsfxsize=2.5truein\epsffile{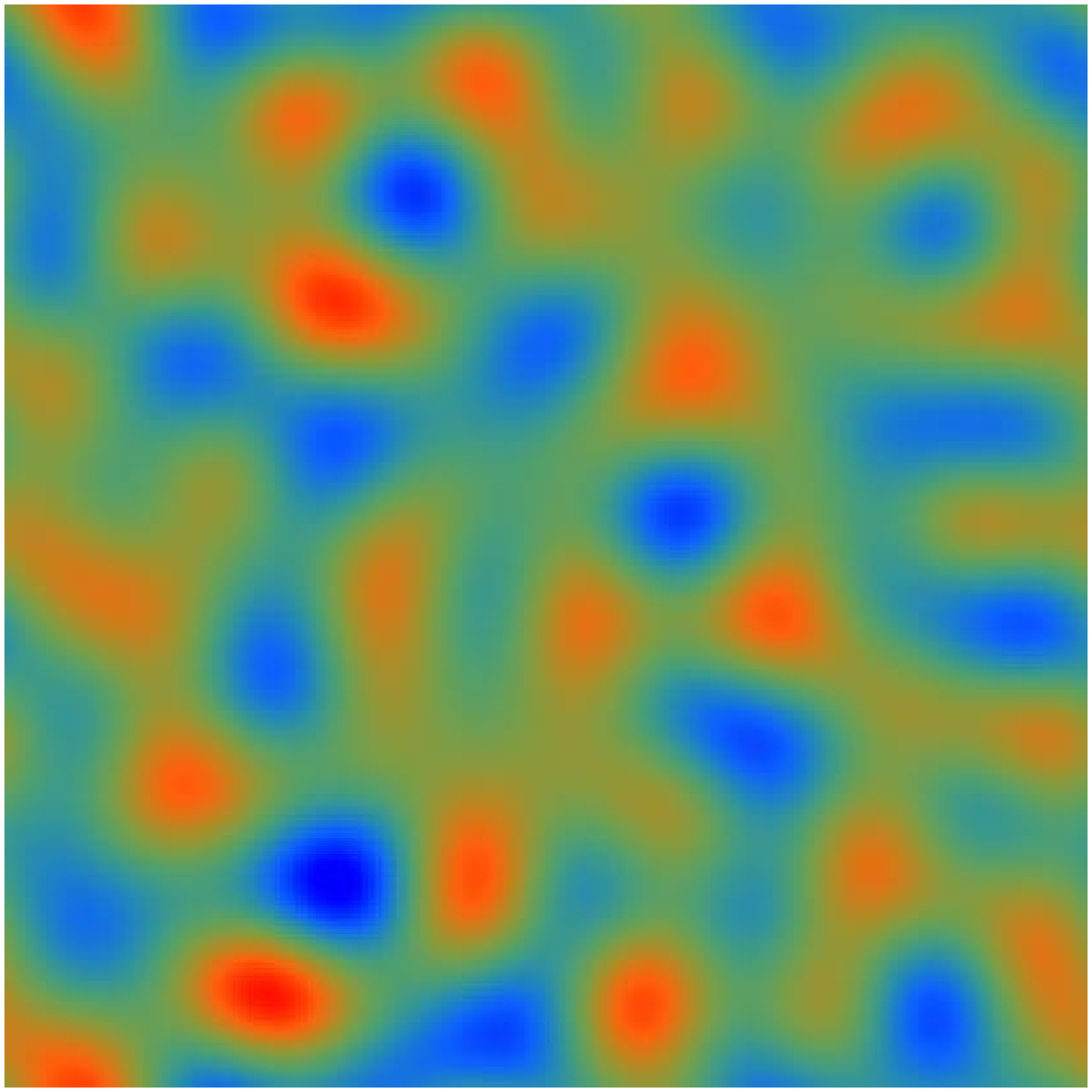}}
\caption{Large-angle ($l<100$) lensing $B$-polarization field (top) and
the reconstructed $B$-polarization field from the small angle $EB$ deflection
estimator and the observed $E$-field.   Detector noise appropriate for
the reference experiment has been added to this $25^\circ \times 25^\circ$
patch.  Reconstruction techniques can help separate the gravitational 
wave and lensing $B$-modes.}
\label{fig:unlens}
\end{figure}

\subsection{Dark Energy and the ISW Effect}

The integrated Sachs-Wolfe (ISW) effect from the differential redshift
due to the decay in the gravitational potential 
is extremely sensitive to the background properties of the
dark energy (\cite{CobDodFri97} 1997; \cite{CalDavSte98} 1998) and provides a unique
handle on its clustering properties (\cite{Hu98} 1998). 
The latter can potentially 
test the scalar-field hypothesis for its nature. 
Unfortunately the ISW effect is buried under the larger primary anisotropy 
and can best be isolated through cross correlation with other large-scale
tracers of the gravitational potential.  The deflection field
of CMB lensing provides a perfect candidate for cross correlation
(\cite{SelZal98} 1998, \cite{GolSpe99} 1999, \cite{Hu01c} 2001c).  
In a flat universe, detection of 
any cross-correlation at all represents an essentially direct
detection of the dark energy.

Since the cross correlation effect is confined to multipoles $L \lsim 100$,
the $\Theta\Theta$ estimator can itself reach the cosmic variance limit
for detection.  What polarization provides is 4 other nearly independent
probes of the cross-correlation.  
Furthermore the polarization estimators each contain enough signal-to-noise
to reconstruct the deflection field with independent sets of multipoles
in the polarization field (see \cite{Hu01c} 2001 for the analogous
technique using the temperature field).
Since the signal is weak and the
importance of understanding the particle properties of the dark energy
great, the ability to make these consistency tests is an important asset.

In Fig.~{\ref{fig:isw}}, we compare the ability of the Planck experiment
$(f_{\rm sky}=0.65)$ and the reference experiment $(f_{\rm sky}=1)$ 
to measure the deflection-ISW cross correlation $C_L^{\Theta d}$.

\subsection{Gravitational Waves and $B$-modes}

The $B$-mode polarization produced by gravitational waves offers what is
perhaps our most direct window on the early universe
(\cite{KamKosSte97} 1997; \cite{ZalSel97} 1997).  
Under the inflationary paradigm, its amplitude determines the
energy scale of inflation.
In this context, gravitational lensing, which also generates
$B$-modes acts as a contaminant.  To constrain inflationary 
energy scales below $10^{16}$ GeV, removal of
the lensing contaminant, either statistical or direct, will be required
(\cite{Hu01c} 2001c).
Fortunately the converse problem does apply:
since the $B$-modes used to reconstruct the deflection fields reside
in the arcminute regime, even much larger amplitude
gravitational wave $B$-modes at degree scales do not contaminate
mass reconstruction.  This fact allows the possibility of direct
removal of the lensing $B$-modes at degree scales.

We save detailed exploration of $B$-mode decontamination to a future
work and here simply show that the mass reconstruction from the
small-scale $B$-polarization itself has sufficient
signal-to-noise to make the procedure feasible.  Decontamination is
not feasible through the $\Theta\Theta$ estimator since most of
the $B$-modes on degree scales arise from fluctuations in the
deflection field with $L \gsim 200$.  

Consider the $E$-modes of the observed, i.e. {\it lensed} and including
detector noise, polarization field and construct the Stokes 
parameters $Q(\bn)$ and $U(\bn)$ from it alone via Eqn.~(\ref{eqn:multipoles}).
Use the reconstructed deflection field $d_{EB}(\bn)$ to artificially
lens the distribution.  The $B$-field of the resulting polarization 
is an estimator of the $B$-field from lensing that is independent of
the true $B$-field on large scales.  
In Fig.~\ref{fig:unlens}, we show the true $B$-field from lensing,
low pass filtered to $l<100$,
compared with the reconstructed $B$-field for the reference experiment on
a $25^\circ \times 25^\circ$ field.
 
\section{Discussion}
\label{sec:discussion}

Based on the induced correlation between the $E$ and $B$ modes,
the lensing of CMB polarization offers the opportunity to reconstruct
the mass distribution in projection on scales corresponding to
$0.002 < k < 0.2$ in $h$ Mpc$^{-1}$.  Compared with a similar reconstruction
from the temperature field, polarization allows for an order of magnitude
extension to smaller scales.  These small scales can correspondingly
be probed with the smaller degree scale fields of view that 
are more typical for planned polarization studies.  Moreover mass reconstruction
is only sensitive to the arcminute scale correlations in the polarization
field and does not require a true map over the full field.
This additional range does not come at the expense of higher 
resolution requirements: the signal-to-noise saturates at the several arcminute 
scale corresponding
to $l \approx 2000$.   Due to the smaller absolute scale of the
signal, polarization studies
do require much more sensitive detectors, with the $EB$ estimator
surpassing the $\Theta\Theta$ estimator for $\Delta_P \lsim 8\mu$K-arcmin
for $L \sim 300$.  Sensitivity and control over foregrounds
and systematics are issues that any polarization based study must 
address, especially those searching for the gravitational wave
imprint from inflation.

Mass reconstruction from CMB polarization can provide measurements 
of the matter power spectrum over a wide range of scales 
that are entirely free of assumptions of how the luminous matter traces the 
mass (or bias) 
and the distribution of lensing sources, as well as largely free
of non-linear corrections.   It complements cosmic shear studies by
providing the deepest two-dimensional mass maps possible 
to anchor tomographic studies of the evolution of structure.  
It extends the shear-based lensing studies
to near horizon-sized structures and therefore provides the opportunity
to study the dark energy in its cross correlation with the ISW effect
in the temperature field.  Finally, since reconstruction only requires
information from fine-scale correlations in the polarization field, it may
be used to remove the lensing $B$-modes on large-scales from any potential
gravitational wave signal.  These potential scientific returns may help
justify the great experimental effort that will be required to map the
CMB polarization field.

\smallskip
{\it Acknowledgments:} We thank Bruce Winstein and Matias Zaldarriaga
for useful discussions.  This work was supported by
NASA NAG5-10840 and the DOE OJI program.

\end{document}